\documentclass[preprint]{vgtc}

\ifpdf%
  \pdfoutput=1\relax
  \usepackage{graphicx}
  \DeclareGraphicsExtensions{.pdf,.png,.jpg,.jpeg}
\else%
  \usepackage{graphicx}
  \DeclareGraphicsExtensions{.eps}
\fi%

\graphicspath{{figures/}{pictures/}{images/}{./}}

\usepackage{xcolor}
\usepackage{microtype}
\PassOptionsToPackage{warn}{textcomp}
\usepackage{textcomp}
\usepackage{amsmath}
\usepackage{amssymb}
\usepackage{mathptmx}
\usepackage{times}

\usepackage{cite}
\usepackage{tabu}
\usepackage{booktabs}
\usepackage{color}
\usepackage{subfigure}
\usepackage{datetime}
\usepackage{enumitem}
\usepackage{hyperref}

\AtBeginDocument{%
}

\hypersetup{
    colorlinks=false,
    pdfborder={0 0 0},
    pdfborderstyle={/S/U/W 0},
}

\newcommand{\mynoenumitemtopsep}{-5pt}

\renewcommand{\vec}[1]{\mathbf{#1}}
\newcommand{\mat}[1]{\mathrm{#1}}
\newcommand{\norm}[1]{\left\lVert#1\right\rVert}

\newcommand{\droplet}{\mathcal{D}}
\newcommand{\refframe}{\mathcal{F}}
\newcommand{\pathline}{\mathcal{L}}

\newcommand{\dropcol}{Colliding Drops}

\onlineid{1075}
\vgtccategory{Research}
\vgtcinsertpkg
\ieeedoi{10.1109/VIS54862.2022.00036}

\title{Droplet-Local Line Integration for Multiphase Flow}

\author{Alexander Straub\thanks{e-mail: alexander.straub@visus.uni-stuttgart.de}\\ %
  \scriptsize University of Stuttgart %
\and Sebastian Boblest\thanks{e-mail: sebastian.boblest@visus.uni-stuttgart.de}\\ %
  \scriptsize University of Stuttgart %
\and Grzegorz K. Karch\thanks{e-mail: grzegorz.karch@visus.uni-stuttgart.de}\\ %
  \scriptsize University of Stuttgart %
\and Filip Sadlo\thanks{e-mail: sadlo@uni-heidelberg.de}\\ %
  \scriptsize Heidelberg University %
\and Thomas Ertl\thanks{e-mail: thomas.ertl@vis.uni-stuttgart.de}\\ %
  \scriptsize University of Stuttgart}

\teaser{
  \centering%
  \def\imagewidth{0.18}%
  \def\clipTB{5}%
  \hfill%
  \subfigure[]{\includegraphics[width=\imagewidth\linewidth,trim=0 \clipTB cm 0 \clipTB cm,clip]{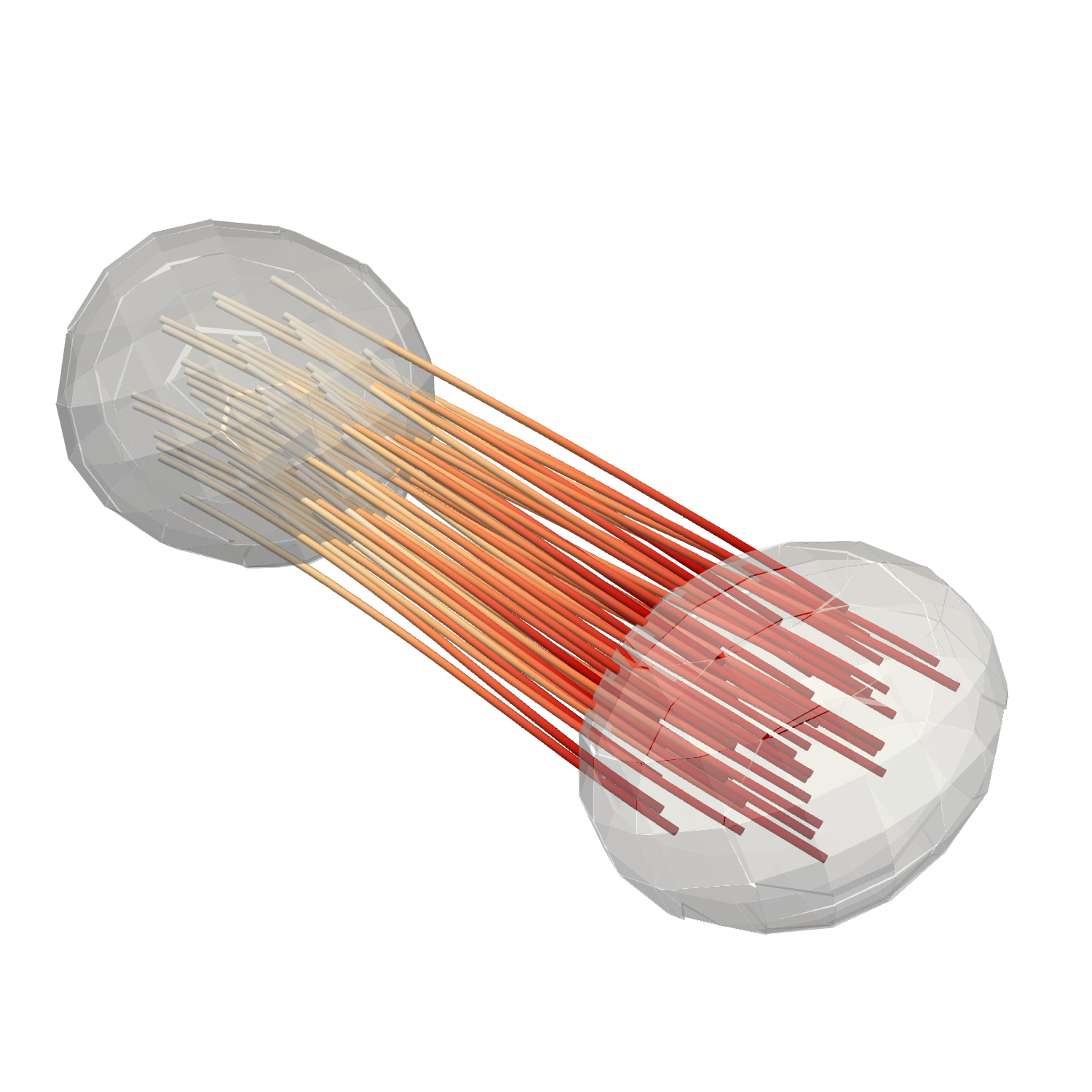}%
    \label{fig:dropcol-500-translation-orig}}%
  \hfill%
  \subfigure[]{\includegraphics[width=\imagewidth\linewidth,trim=0 \clipTB cm 0 \clipTB cm,clip]{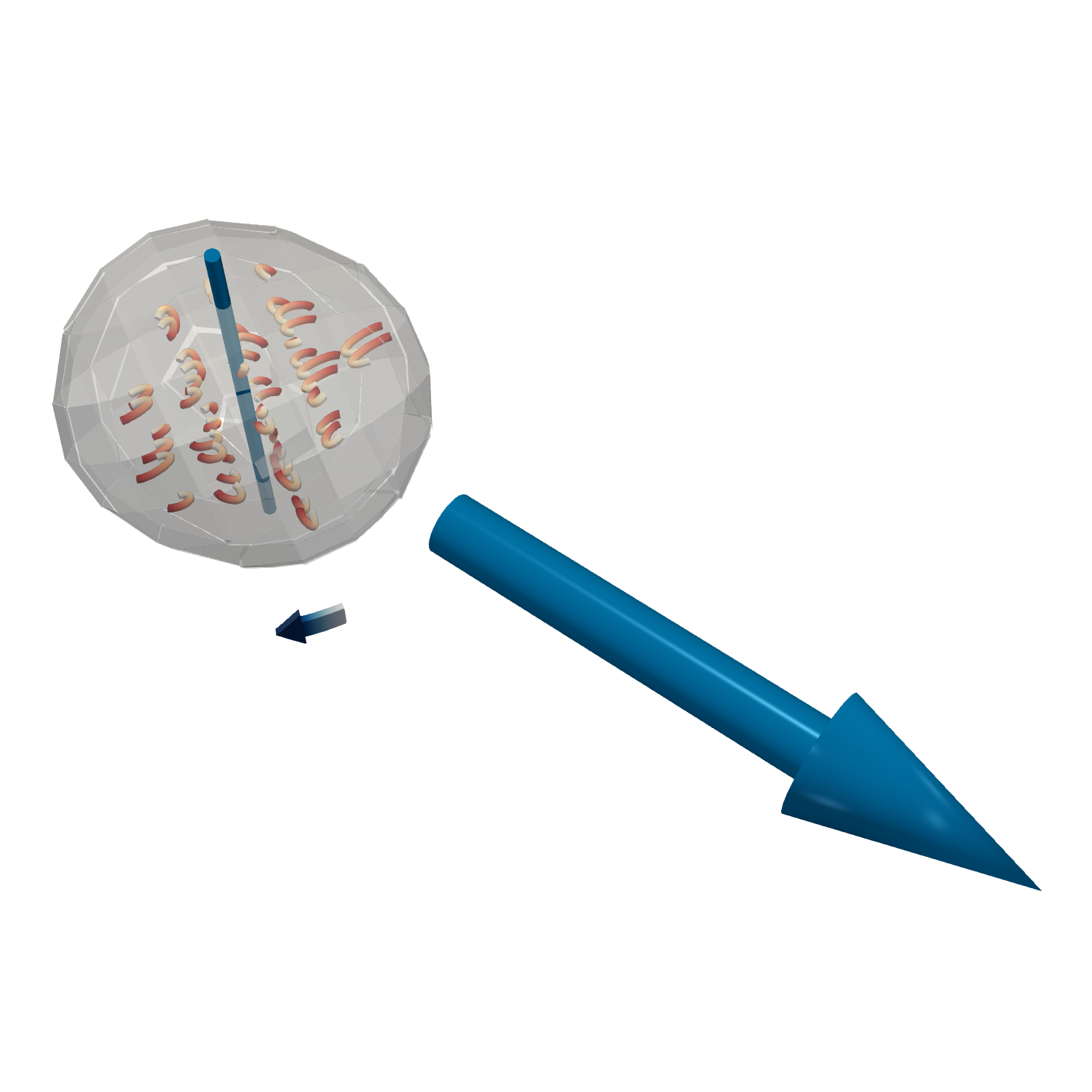}%
    \label{fig:dropcol-500-translation}}%
  \hfill%
  \subfigure[]{\includegraphics[width=\imagewidth\linewidth,trim=0 \clipTB cm 0 \clipTB cm,clip]{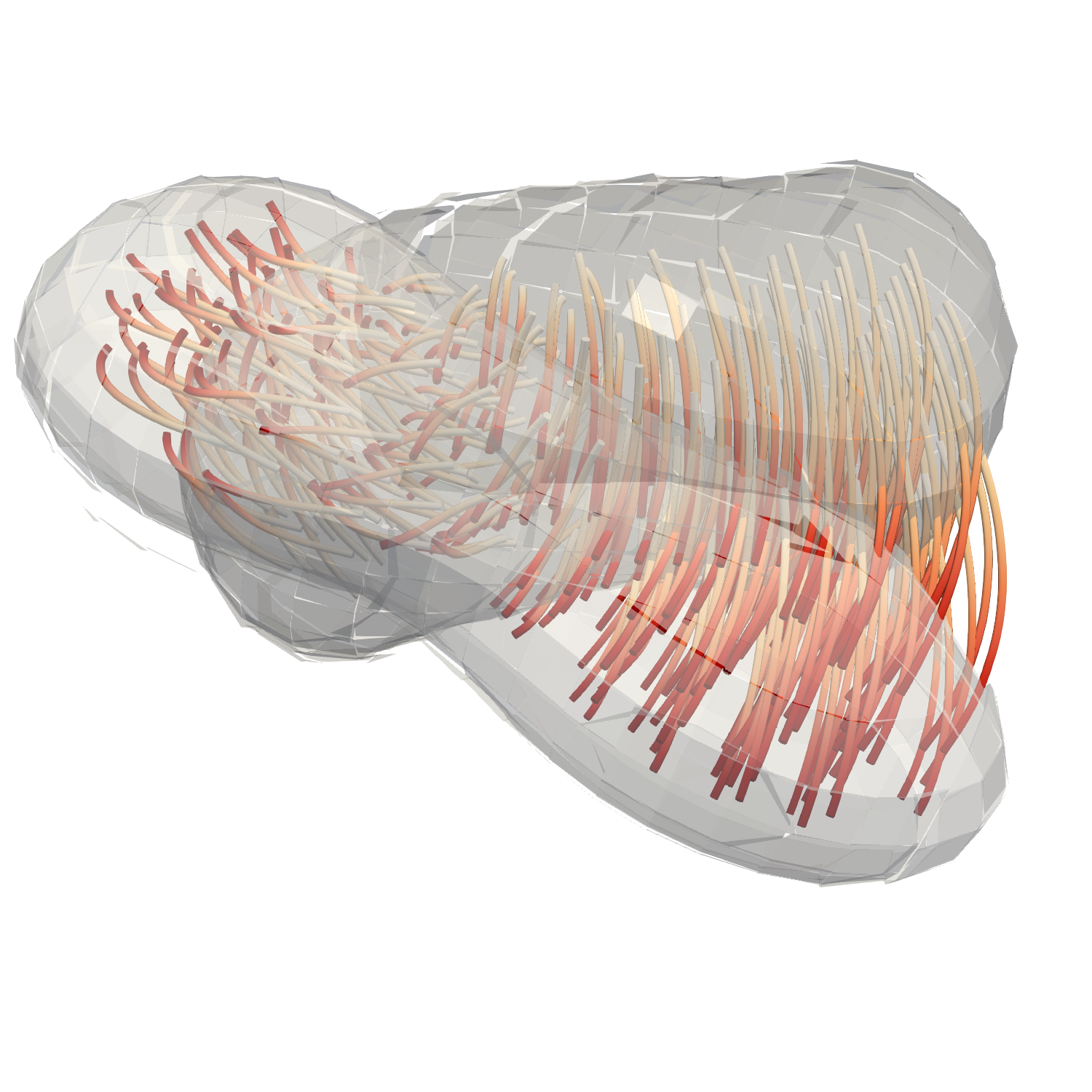}%
    \label{fig:dropcol-500-rotation-orig}}%
  \hfill%
  \subfigure[]{\includegraphics[width=\imagewidth\linewidth,trim=0 \clipTB cm 0 \clipTB cm,clip]{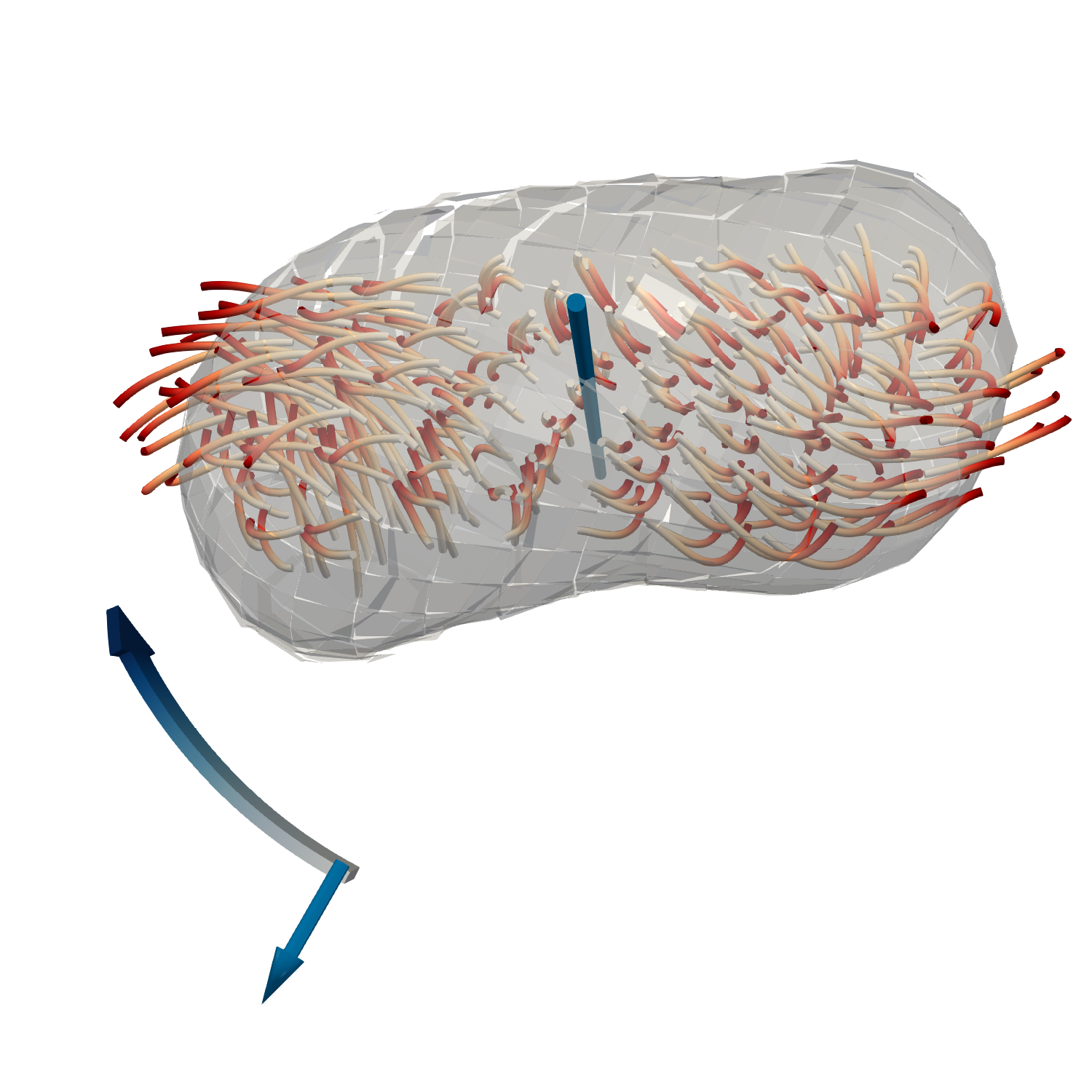}%
    \label{fig:dropcol-500-rotation}}%
  \hfill%
  \caption{%
    Pathlines for two examples from the~\dropcol~dataset.
    Droplets exhibit large translational~\subref{fig:dropcol-500-translation-orig} and rotational~\subref{fig:dropcol-500-rotation-orig} velocity components that overshadow the internal flow.
    \subref{fig:dropcol-500-translation},\subref{fig:dropcol-500-rotation}~Droplet-local flow reveals more detailed flow patterns.
    Glyphs replace removed velocity components: translation (round arrow) and rotation (bent arrow and axis).
    Flow direction from white to red.
  }%
\label{fig:dropcol}%
}

\abstract{%
Line integration of stream-, streak-, and pathlines is widely used and popular for visualizing single-phase flow.
In multiphase flow, i.e., where the fluid consists, e.g., of a liquid and a gaseous phase, these techniques could also provide valuable insights into the internal flow of droplets and ligaments and thus into their dynamics.
However, since such structures tend to act as entities, high translational and rotational velocities often obfuscate their detail.
As a remedy, we present a method for deriving a droplet-local velocity field, using a decomposition of the original velocity field removing translational and rotational velocity parts, and adapt path- and streaklines.
Generally, the resulting integral lines are thus shorter and less tangled, which simplifies their analysis.
We demonstrate and discuss the utility of our approach on droplets in two-phase flow data and visualize the removed velocity parts employing glyphs for context.
}

\CCScatlist{
  \CCScatTwelve{Human-centered computing}{Visu\-al\-iza\-tion}{Visu\-al\-iza\-tion application domains}{Scientific visualization};
}

\begin{document}

\firstsection{Introduction}
\label{sec:introduction}
\maketitle

Simulation and investigation of multiphase flow are driving topics of today's computational fluid dynamics research.
Ap\-pli\-ca\-tions include the development of combustion engines, spray cooling systems for food storage, as well as improving models for weather forecasts.
In this area, interesting topics include the study of droplet interactions, but also the analysis of flow dynamics within individual droplets.
In helping domain experts understand the results of their simulations, visualization plays an important role.

One of the most widely used techniques for flow visualization are line integration methods.
Stream-, streak-, and pathlines were originally developed for single-phase flow.
While their application to two-dimensional data is straightforward, application to three-dimensional flow already poses challenges such as occlusion.
To this end, much work has been published to address the reduction of visual clutter.
Notable techniques include line placement for sparse seeding of integral lines~\cite{Turk1996,McLoughlin2013}, as few lines usually suffice to describe the overall flow behavior.
This can be extended to view-dependent visualization~\cite{Marchesin2010}.
Another set of methods employs opacity adjustment.
Here, an importance criterion is mapped to the opacity of the integral lines, blending out ``irrelevant'' parts~\cite{Guenther2014,Guenther2017}.
A different concept is to replace flow lines with glyphs, where one can simultaneously visualize different aspects of the flow~\cite{Brambilla2013} or reduce overdraw and highlight regions~\cite{Hlawatsch2014}.
While our method reduces visual clutter, we do not aim to replace the aforementioned techniques but provide an approach that can be employed complementarily.

Vector field decomposition has often been investigated in flow analysis and visualization.
For example, by employing the Helmholtz--Hodge decomposition~\cite{Bhatia2014} or the Green function~\cite{Li2006}, the vector field can be decomposed into an incompressible, irrotational, and harmonic part.
In our work, the components of interest are translation, rotation, and a ``droplet-local'' term.
Contrary to Karch et al.~\cite{Karch2018}, who extract the rotation of a droplet from its shape by using principal component analysis, we use the underlying velocity field.
This way, angular velocity is computed directly from inertia and angular momentum.
This bares the advantage of a more robust computation for spherical and oscillating droplets.
Choosing appropriate frames of reference has been discussed before in order to visualize vortices~\cite{Wiebel2005,Guenther2017a}.
Hadwiger et al.~\cite{Hadwiger2019} bridged the gap between a single, global frame of reference and calculating a frame of reference for each individual position.
Our method, on the other hand, proposes a single frame of reference for each individual droplet.
Additionally, we employ glyphs for visualizing these frames of reference.

To the best of our knowledge, no research has been conducted explicitly on integral lines in three-dimensional multiphase flow.
Addressing this lack, our paper aims at providing the means necessary for domain experts to visually analyze such flow.
This entails gaining an overview for fast-moving and strongly rotating droplets, as well as allowing domain experts to study the interdependence between internal flow and forces acting on droplet interfaces, e.g., on droplet breakup.
To this end, the contributions of this work are:
\begin{itemize}[noitemsep,topsep=\mynoenumitemtopsep]
  \item a physically interpretable droplet-local velocity field,
  \item the adaptation and generalization of streak- and pathlines, and
  \item glyphs conveying the frames of reference.
\end{itemize}
\vspace*{1.3ex}
Please note that the research presented in this paper originates from a collaborative interdisciplinary project on droplet interfaces~\cite{DROPIT} and that its idea and preliminary results were already shortly presented, among other results, in an overview article~\cite{Straub2020} and at a project workshop~\cite{Straub2019} as an extended abstract.
In this article, we now present the technical details of our method, together with the following contributions that have been added:
\begin{itemize}[noitemsep,topsep=\mynoenumitemtopsep]
  \item calculation of droplet rotation based on rigid body mechanics,
  \item glyphs conveying the frames of reference, and
  \item discussion and implementation of a static frame of reference.
\end{itemize}
\vspace*{1.3ex}

\section{Droplet-Local Velocity}
\label{sec:velocities}

High translational and rotational velocity components tend to obfuscate details of the droplet-local flow.
For example, following a collision, a droplet may be separated and ejected at a relatively high velocity.
Thus, its pathlines would be almost straight lines and hence overshadow local detail.
Additionally, rotational motion can lead to visual clutter, overdrawing other integral lines.
To provide a droplet-local view, we, therefore, free the original velocity field from translational and rotational parts.
This modified velocity field can then be used for visualization, using techniques such as stream-, streak- or pathlines, as well as other flow visualization methods.

To obtain such a reduced velocity field~$\vec{\tilde{u}}(\vec{x}, t)$, the first step is to decompose the original velocity field~$\vec{u}(\vec{x}, t)$ into parts:
\begin{equation}
  \vec{u} = \vec{u}_{c} + \vec{u}_{\omega} + \vec{\tilde{u}},
\label{eq:decomposition}
\end{equation}
with droplet translation~$\vec{u}_{c}$, velocity~$\vec{u}_{\omega}$~representing droplet rotation, and ``droplet-local velocity part''~$\vec{\tilde{u}}$.
We thus gain our droplet-local velocity field as~$\vec{\tilde{u}} = \vec{u} - \vec{u}_{c} - \vec{u}_{\omega}$.
This elimination of translational and rotational velocity parts can be intuitively described as a change of reference.
Instead of a global view of the simulation, each droplet is assigned an observer that moves with the droplet's translational velocity and rotates at the droplet's angular velocity about its axis.
As a consequence, certain physical properties are maintained, such as the conservation of mass and volume in case of incompressible flow, which increases (physical) interpretability.

Note that in contrast to other decomposition schemes, such as Helmholtz-Hodge, the droplet-local velocity is defined for the whole droplet as if it was a rigid body.
For this, a droplet~$\droplet$ is defined as a connected region of cells that contains the same fluid phase and which is separated from other droplets by at least one boundary layer of ``empty'' cells~(cf.~\autoref{sec:results} for a more detailed definition).

As such, a droplet can generally be of arbitrary shape and size.

\subsection{Droplet Translation and Rotation}

The translational velocity of a droplet~$\droplet$ is its linear velocity
\begin{equation}
  \vec{u}_c = \frac{\int_\droplet{\rho \cdot \vec{u}~\mathrm{d}V}}{\int_\droplet{\rho~\mathrm{d}V}}
    \approx \frac{\sum_i \rho_i V_i \cdot \vec{u}_i}{\sum_i \rho_i V_i},
\end{equation}
with original velocity~$\vec{u}$, density~$\rho$, and volume~$V$.

Contrary to the translational velocity, which is given for the whole droplet, the rotational velocity depends on the location relative to the axis of rotation.
This axis goes through the center of mass
\begin{equation}
  \vec{r}_c = \frac{\int_\droplet{\rho \cdot \vec{r}~\mathrm{d}V}}{\int_\droplet{\rho~\mathrm{d}V}}
    \approx \frac{\sum_i \rho_i V_i \cdot \vec{r}_i}{\sum_i \rho_i V_i},
\end{equation}%
integrated over the positions~$\vec{r}$.
For the calculation of the angular velocity, positions have to be considered relative to the center of mass, hence~$\vec{r}' = \vec{r} - \vec{r}_c$.
Angular velocity~$\vec{\omega}$~is defined as 
\begin{equation}
  \vec{\omega} = \mat{I}^{-1} \vec{L},
\end{equation}
for which first the inertia tensor~$\mat{I}$ is constructed for a droplet.
Its components are given as
\begin{align}
  I_{qr} = \int_\droplet{\rho \cdot \left( \norm{\vec{r}'}^2 \delta_{qr} - (\vec{r}')_q (\vec{r}')_r \right)\,\mathrm{d}V} \\
    \approx \sum_i \rho_i V_i \cdot \left( \norm{\vec{r}_i'}^2 \delta_{qr} - (\vec{r}_i')_q (\vec{r}_i')_r \right),
\end{align}%
where~$\delta_{qr}$ is the Kronecker delta, and~$(\vec{r}')_n$ is the~$n$-th component of~$\vec{r}'$.
Next, the total angular momentum~$\vec{L}$ for the rotation of the droplet around its axis is calculated as
\begin{equation}
  \vec{L} = \int_\droplet{\rho \cdot \left( \vec{r}' \times \vec{u} \right)\,\mathrm{d}V} \approx \sum_i \rho_i V_i \cdot \left( \mathbf{r}'_i \times \mathbf{u}_i \right).
\end{equation}%
The resulting rotational velocity part is then
\begin{equation}
  \vec{u}_{\omega} = \vec{\omega} \times \vec{r}'.
\end{equation}

\section{Line Integration}
\label{sec:integration}

For static visualization, the droplet-local vector field~$\vec{\tilde{u}}$ can be used directly for streamline computation.
However, to apply streak- and pathlines for visualization of time-dependent flow, these two concepts need to be adapted.
This is because (1)~the fluid phase moves over time, and (2)~we have a different frame of reference for each droplet.
Additionally, for streaklines, this means that (3)~the seed itself has to move in the droplet's frame of reference.

\subsection{Pathlines}
\label{sec:integration-pathlines}

\begin{figure}
  \centering%
  \def\illuheight{2.3cm}%
  \subfigure[]{\includegraphics[height=\illuheight]{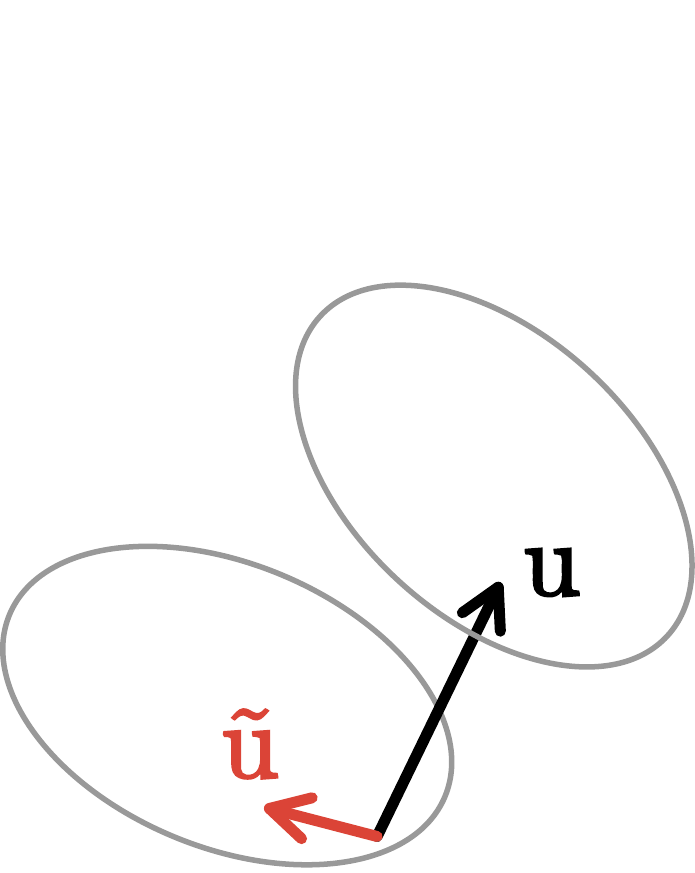}\label{fig:pathlines-explain-1}}%
  \hfill%
  \subfigure[]{\includegraphics[height=\illuheight]{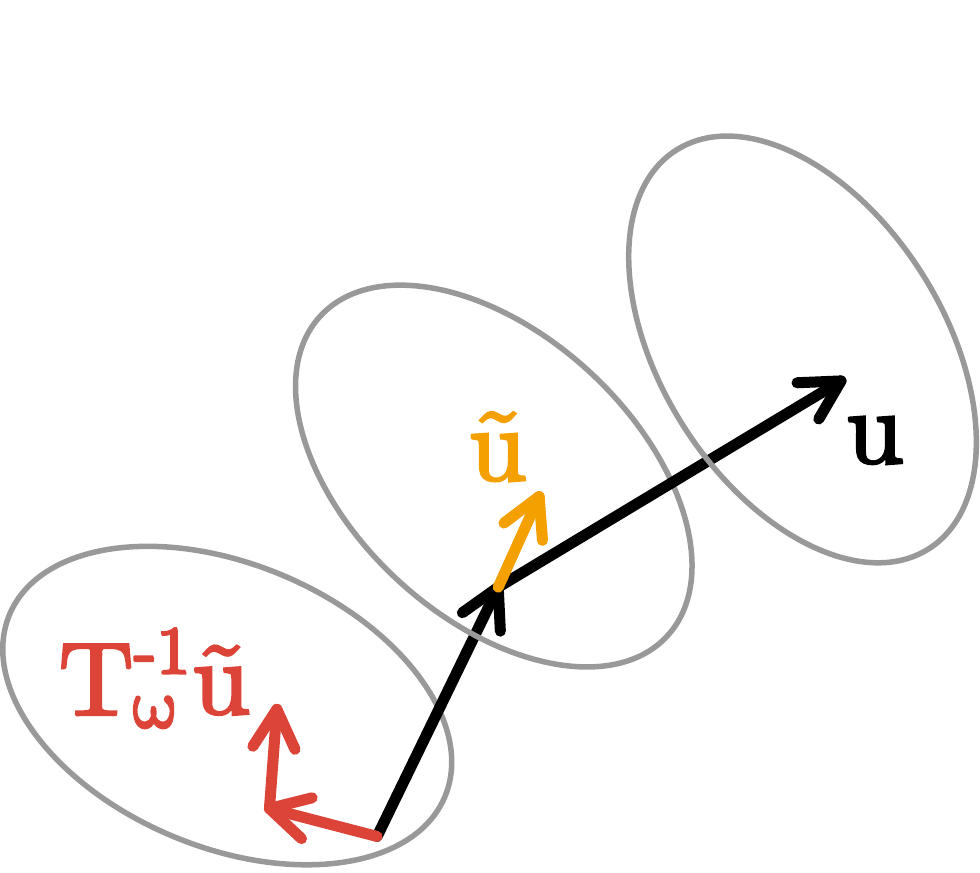}\label{fig:pathlines-explain-2}}%
  \hfill%
  \subfigure[]{\includegraphics[height=\illuheight]{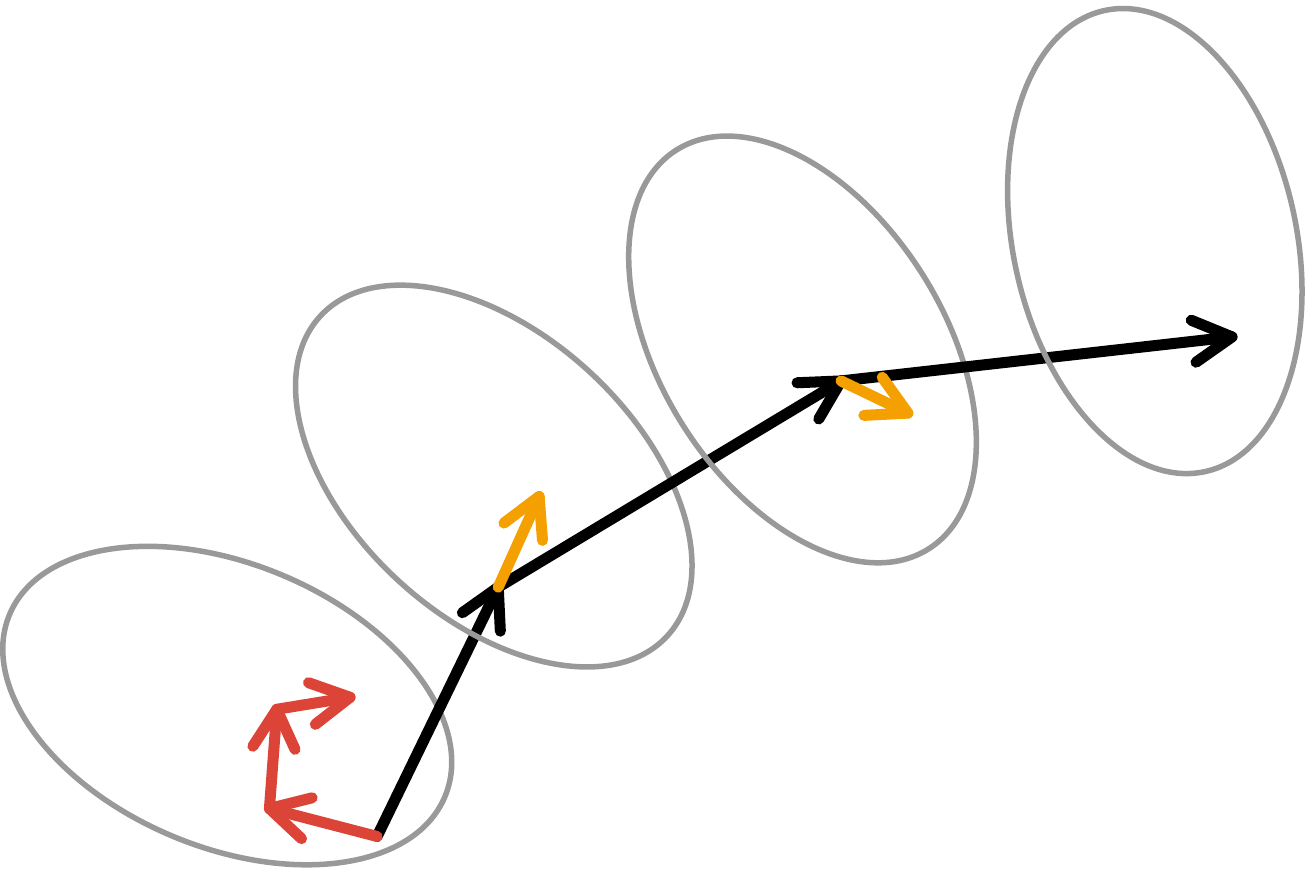}\label{fig:pathlines-explain-3}}%
  \caption{%
    Time series of the integration within the original velocity field~$\vec{u}$ (black), and simultaneous integration within the droplet-local velocity field~$\left( \mat{T}_{\omega}^{-1}~\vec{\tilde{u}} \right)$~(red).
    \subref{fig:pathlines-explain-1}~Advection of~$\vec{x}(t_0)$ in the original velocity field~$\vec{u}$~(black arrow), and in the droplet-local velocity field~$\vec{\tilde{u}}$~(red arrow).
    \subref{fig:pathlines-explain-2}--\subref{fig:pathlines-explain-3}~In the subsequent time steps, the droplet-local velocities~$\vec{\tilde{u}}$~(orange arrows) are sampled at the positions~$\vec{x}$ of the advected original particle and used to advect the droplet-local particle po\-sitions~$\vec{\tilde{x}}$ using the transformed velocity field~$\left( \mat{T}_{\omega}^{-1}~\vec{\tilde{u}} \right)$~(red arrows).
  }
  \label{fig:pathlines-explain}
\end{figure}

For the computation of droplet-local pathlines, the idea is to integrate pathlines~$\smash{\pathline(t):=\lbrace \vec{x}(t) \rbrace}$ within the original velocity field~$\vec{u}(\vec{x}(t), t)$, and use their positions to sample the droplet-local velocity field~$\vec{\tilde{u}}(\vec{x}(t), t)$.
The extracted droplet-local velocity is then used to integrate the droplet-local pathline~$\smash{\widetilde{\pathline}(t):=\lbrace \vec{\tilde{x}}(t) \rbrace}$.
This is illustrated in~\autoref{fig:pathlines-explain}, where the advection of the original particle is visualized by black arrows and the droplet-local advection by red arrows.
Due to the sampling along the pathline in the original field, sample positions are within the correct phase.
Using the droplet-local field for the corresponding time step, we obtain the droplet-local velocity at the sample positions.
This velocity is shown as orange arrows in the figure.
Transformed back, it is then used to advect the droplet-local particle position~(red arrows).

Mathematically, pathlines are expressed as initial value problem:
\begin{equation}
  \frac{\mathrm{d} \vec{x}(t)}{\mathrm{d} t} = \vec{u}(\vec{x}(t), t),~~~~\vec{x}(t_0) = \vec{x}_0.
\label{eq:pathline}
\end{equation}
In our method, they are used for generating the pathline in the original vector field.
This pathline is not visualized, but its calculated position~$\vec{x}(t)$ is used to sample the droplet-local velocity:
\begin{equation}
  \frac{\mathrm{d} \vec{\tilde{x}}(t)}{\mathrm{d} t} = \mat{T}_{\omega}^{-1}~\vec{\tilde{u}}(\vec{x}(t), t),~~~~\vec{\tilde{x}}(t_0) = \vec{x}_0.
\label{eq:droplet-local-pathline}
\end{equation}
Here, the droplet-local velocity field~$\vec{\tilde{u}}(\vec{x}(t), t)$ is used for the calculation of~$\vec{\tilde{x}}(t)$.
The inverse rotation~$\smash{\mat{T}_{\omega}^{-1}}$ is used to transform the velocity back to the droplet's original coordinate system at seeding time~$t_0$.
For each of the three axes~$\vec{e}_i$ of the droplet's local, co-rotating frame, we have
\begin{equation}
  \frac{\mathrm{d} \vec{e}_i}{\mathrm{d} t} = \omega\times\vec{e}_i,
\label{eq:rotating-frame-1}
\end{equation}
\begin{equation}
  \vec{e}_i(t) = \int_{t_0}^{t}\omega(t')\times\vec{e}_i(t')~\text{d} t' + \vec{e}_i(t_0).
\label{eq:rotating-frame-2}
\end{equation}
Bases~$\smash{\mat{E}(t) = \left< \vec{e}_0(t)~\vec{e}_1(t)~\vec{e}_2(t) \right>}$ and~$\smash{\mat{E}(t_0) = \left< \vec{e}_0(t_0)~\vec{e}_1(t_0)~\vec{e}_2(t_0) \right>}$ can then be used to calculate the rotation matrix~$\smash{\mat{T}_{\omega}^{-1}}$, which transforms~$\mat{E}(t)$ to $\mat{E}(t_0)$.
This back-trans\-for\-ma\-tion can be observed in~\autoref{fig:pathlines-explain}.
Note that the sampled~(orange arrows) and the droplet-local velocities~(red arrows) are rotated by~$\mat{T}_{\omega}$.

\begin{figure}[!t]%
  \centering%
  \subfigure[]{\includegraphics[width=0.47\linewidth]{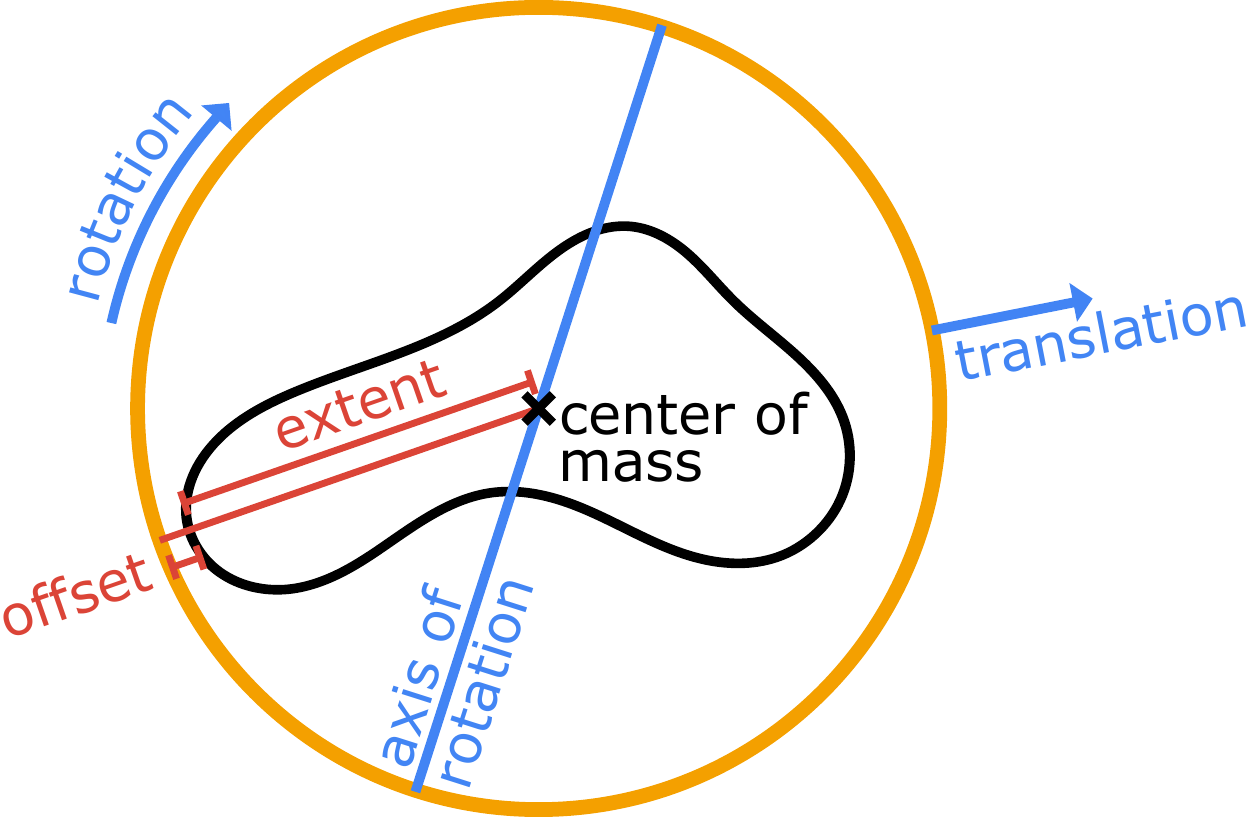}%
    \label{fig:context-illustration-illu}}%
  \hspace{3em}%
  \subfigure[]{\hspace{-2em}\includegraphics[width=0.4\linewidth,trim=0 200px 0 100px,clip=true]{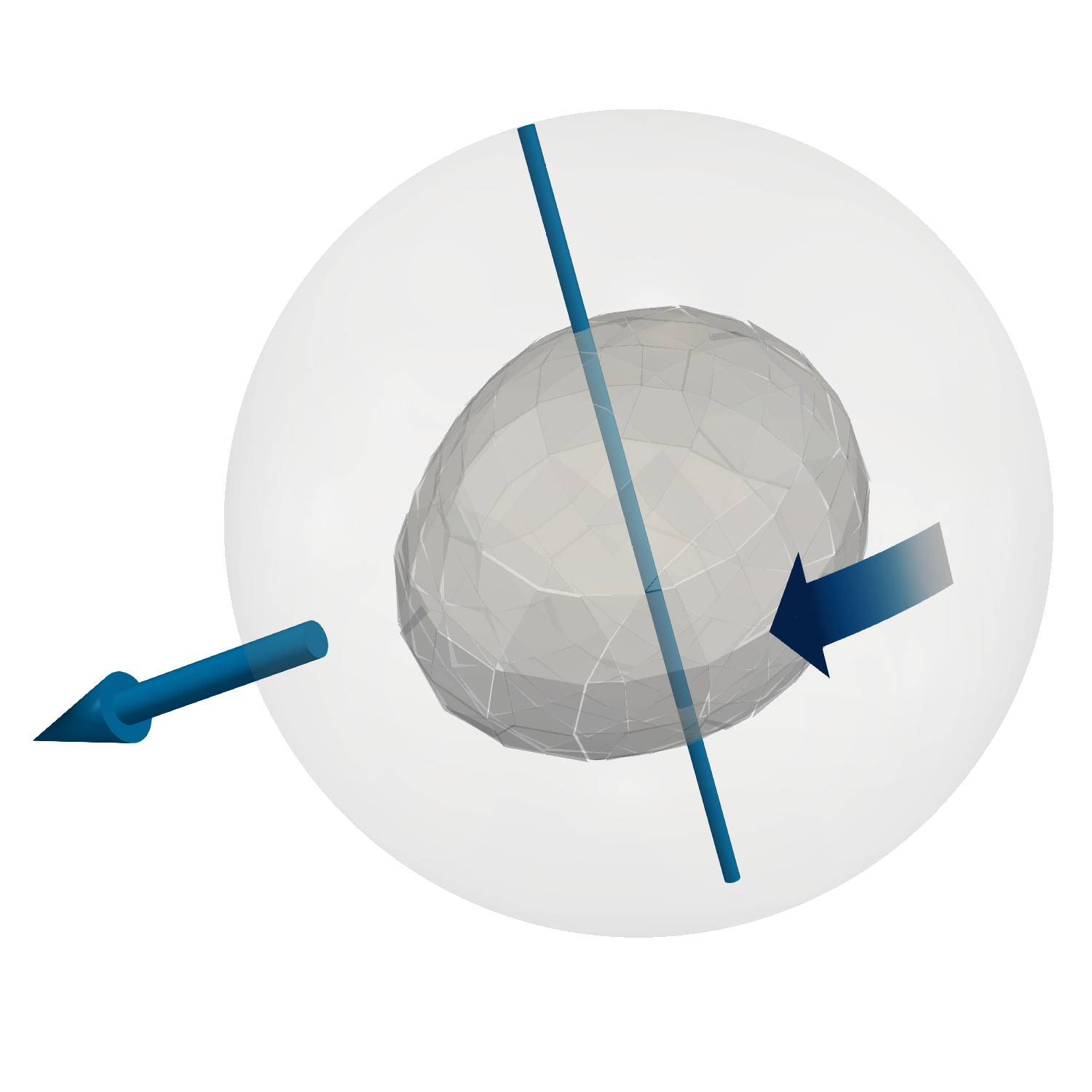}%
    \label{fig:context-illustration-real}}%
  \caption{%
    Glyphs conveying the frame of reference.
    \subref{fig:context-illustration-illu}~Different aspects of information surrounding the droplet~(black).
    A cir\-cum\-sphere~(orange) around the center of mass is constructed for positioning, its radius slightly larger than the droplet~(indicated in red).
    The axis of rotation~(blue, middle) goes through the center of mass, its length the diameter of the sphere.
    Effective rotation~(blue, upper left) is visualized as a bent arrow~(blue) with a corresponding length.
    The arrow visualizing translation~(blue, right) is also attached to the sphere, its length coincides with the actual translation.
    \subref{fig:context-illustration-real}~Example droplet from the~\dropcol~dataset, showing translation~(volumetric blue arrow), axis of rotation~(blue tube), and effective rotation~(curved flat arrow with time additionally encoded from white to blue).
  }%
  \label{fig:context-illustration}%
\end{figure}%

\subsection{Streaklines}
\label{sec:integration-streaklines}

Streaklines can be generated following the same idea as for pathlines.
Again, we use the original velocity field for finding the sample positions, where we interpolate the droplet-local field to integrate the droplet-local streakline.
Similar to generalized streaklines, which were introduced by Wiebel et al.~\cite{Wiebel2007}, the seed has to be moved.
This is, among others, necessary because otherwise, an increasing number of particles would be inserted outside the correct phase.
While Wiebel et al. keep the seed static relative to a feature, we have to keep it static relative to the droplet-local coordinate system.
For this, the seed is translated with the droplet's linear velocity, as well as rotated around its axis of rotation.

\section{Frame of Reference}
\label{sec:context}

Our previously described method inherently uses a dynamic frame of reference~$\refframe_\droplet(t)$.
This means that in every time step, we compute for each droplet a new frame of reference from the instantaneous velocity field.
Thus, the translation and rotation of the droplet, which form the frame of reference, may change over time.

\subsection{Static Frame of Reference}

Additionally, we introduce the use of a static frame of reference, which means that at time~$t_0$, each droplet~$\droplet$ is assigned its frame of reference~$\refframe_\droplet(t=t_0)$.
This frame of reference then does not change over time, i.e.,~$\refframe_\droplet(t):=\refframe_\droplet(t_0)$.
To assign a droplet at time step~$t_n$ its corresponding frame of reference extracted at time~$t_0$, we need to be able to track droplets over time.
Note that this is not necessary for the dynamic frame of reference, as it only depends on the current time step.
However, for the dynamic frame, we need to track droplets over time for our context visualization~(cf.~\autoref{sec:visualization}).
For this, we employ a similar scheme to that of Karch et al.~\cite{Karch2018}, advecting the droplets' centers of mass in both forward and reverse time.
The advected positions are then compared to the droplets in the respective time step.
This way, additionally to finding the corresponding droplets in neighboring time steps, we can also identify droplet collisions~(forward time) and breakups~(reverse time) by counting the number of advected centers of mass corresponding to a droplet at~$t_{n+1}$, and~$t_{n-1}$, respectively.
Note that droplet collision unavoidably forces the integration to stop for the involved droplets in case of a static frame of reference, as the frame of reference of the resulting droplet cannot be determined unambiguously, i.e., it would need to result from two droplets at~$t_{n-1}$.

\subsection{Visualization}
\label{sec:visualization}

To mitigate the loss of information from removing velocity parts~$\vec{u}_c$ and~$\vec{u}_\omega$, we provide context visualization for each droplet, as illustrated in~\autoref{fig:context-illustration}.
Here, glyphs are added to show droplet translation and rotation, respectively.
To convey translation, a simple arrow glyph is employed, whose length coincides with the actual translation of the droplet from~$t_0$ to~$t$.
For its placement, we use a circumsphere around the center of mass of the droplet, with a radius slightly larger than the maximum extent.
We then place the glyph on the sphere in the direction of translation.
Rotation, on the other hand, is depicted twofold---employing a bent arrow glyph as well as showing rotation axes.
The arrow glyph shows the actual rotation of the droplet from time~$t_0$ to~$t$.
It is placed on the surface of the circumsphere and is colored according to integration time.
This way, its temporal information is also accessible when viewed from above.
Further, rotation axes are shown to indicate the orientation of rotation.
These tube glyphs go through the droplet's center of mass and are bounded by the circumsphere.
As for a dynamic frame of reference the droplet's axis of rotation changes over time, we allow visualizing all axes from discrete integration steps~$t_i \in [t_0,t]$ at once.
Mapping color to integration time, the temporal evolution of the frame of reference can thus be observed.
This way, we are also able to find numerical instabilities indicated by strongly varying axes, e.g., in the case of small droplets.
As for static frames of reference, we need to track droplets over time to visualize corresponding \mbox{glyphs} at the position of the droplet at time~$t_0$.
To additionally provide the user with context information independent of camera position, we further provide options to duplicate the arrow glyphs to opposite sides of the droplet.
Please also see the video in the supplemental material for further examples.

\section{Results}
\label{sec:results}

We implemented our method as ParaView~\cite{Ayachit2015} plugin, where droplet extraction, line integration, and context visualization are separate filters.
This modular design allows easy adaptation of our method to different data.
The code can be found on GitHub as part of TPF: \urlstyle{same}\url{https://github.com/UniStuttgart-VISUS/tpf}.

While the presented method can be extended to any kind of data that allows segmentation into distinct droplets or clusters, we focus on datasets that stem from direct numerical simulations (DNS) performed using the computational fluid dynamics~(CFD) solver Free Surface 3D~(FS3D)~\cite{Eisenschmidt2016} for incompressible multiphase flow.
The data is generated and stored per cell on a rectilinear grid, with a velocity field~$\vec{u}(\vec{x},t)$ and a volume of fluid~(VOF)~\cite{Hirt1981} field~$f(\vec{x},t)$.
The latter indicates the ratio of the fluid phase for a cell, i.e., $f = 1$ if the cell contains only liquid and $f = 0$ if the cell is ``empty''.
For values~$0 < f < 1$, the liquid occupies a fraction of the cell and thus contains the fluid interface (droplet surface).
Therefore, we define a droplet as a connected region of cells for which~$f > 0$.
This means that different droplets are separated by at least one layer of empty cells.
Note that for VOF data, the discrete volume for each cell is $V_i = f \cdot V_{cell}$.
Please also note that particles are considered to be inside of the droplet if they are in a cell with~$f > 0$, although the reconstructed interface may indicate otherwise.
In the following, we show the results of our method on the~\dropcol~dataset, where two equally large droplets collide in an off-center head-on collision.
This leads to the formation of a single large structure, which eventually disintegrates into a multitude of smaller droplets.

To show the general usefulness of our method, two different droplets from the~\dropcol~dataset, one with a relatively large translational and one with a relatively large rotational velocity part, are visualized in~\autoref{fig:dropcol}.
In both cases, we can see that these large velocity parts hinder us from interpreting the internal flow directly.
For example, in~\autoref{fig:dropcol-500-rotation}, removing dominant droplet rotation reveals slight rotation within the left and right parts in a counter-clockwise direction.
Adding context visualization to the droplet-local pathlines, we are not only provided with the same information as from the original pathlines but can now observe the intricate droplet-local flow.

\begin{figure}[!t]%
  \centering%
  \def\imagewidth{0.24}%
  \def\clipTB{2}%
  \begin{minipage}{0.05\linewidth}%
    \rotatebox[origin=c]{90}{\footnotesize\hspace{-1.5em} Droplet-local \hspace{3.5em} Original}%
  \end{minipage}%
  \begin{minipage}{0.95\linewidth}%
    \hfill%
    \subfigure[Stream]{\includegraphics[width=\imagewidth\linewidth,trim=0 \clipTB cm 0 \clipTB cm,clip]{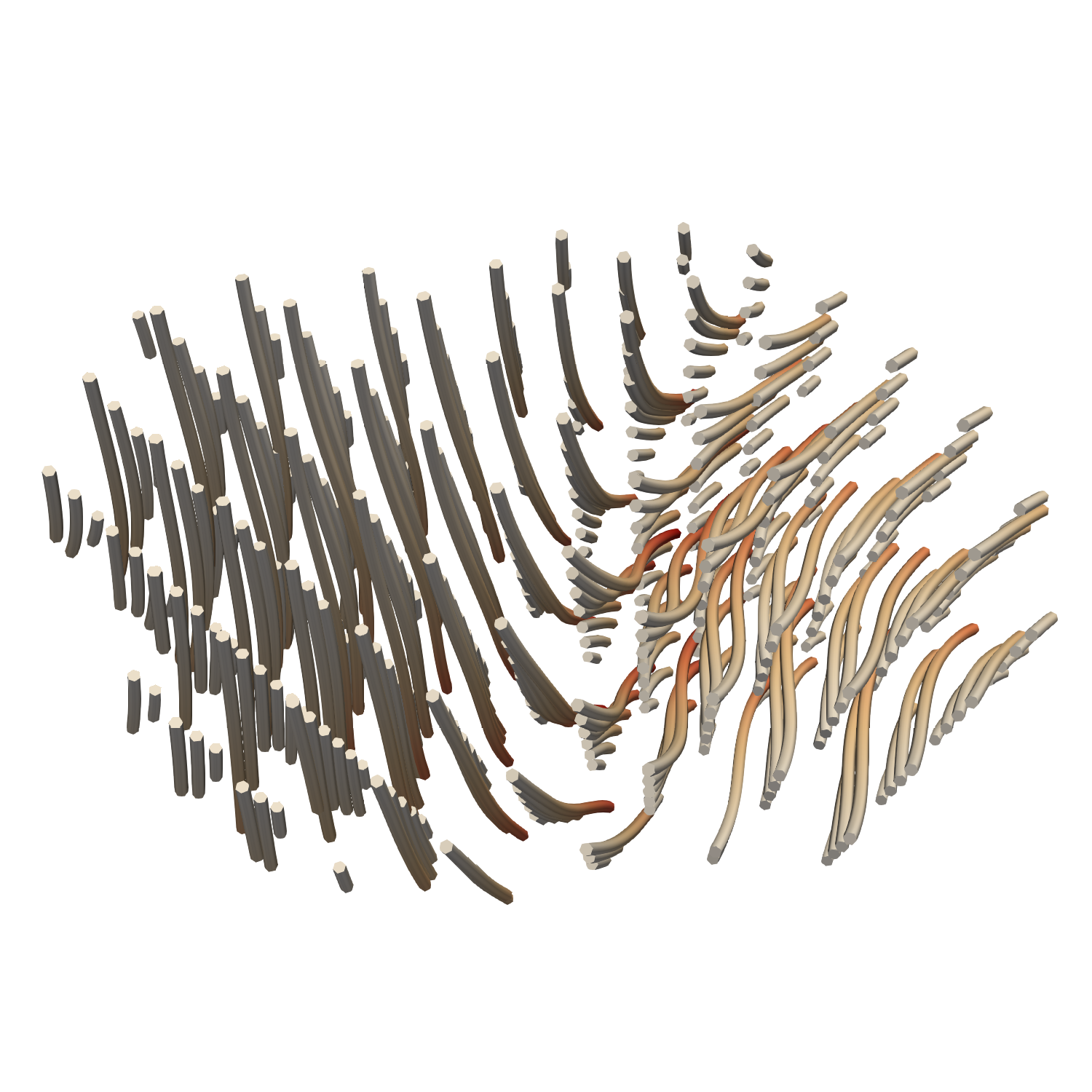}
      \label{fig:dropcol-results-vortex-stream-500-orig}}%
    \hfill%
    \subfigure[Path]{\includegraphics[width=\imagewidth\linewidth,trim=0 \clipTB cm 0 \clipTB cm,clip]{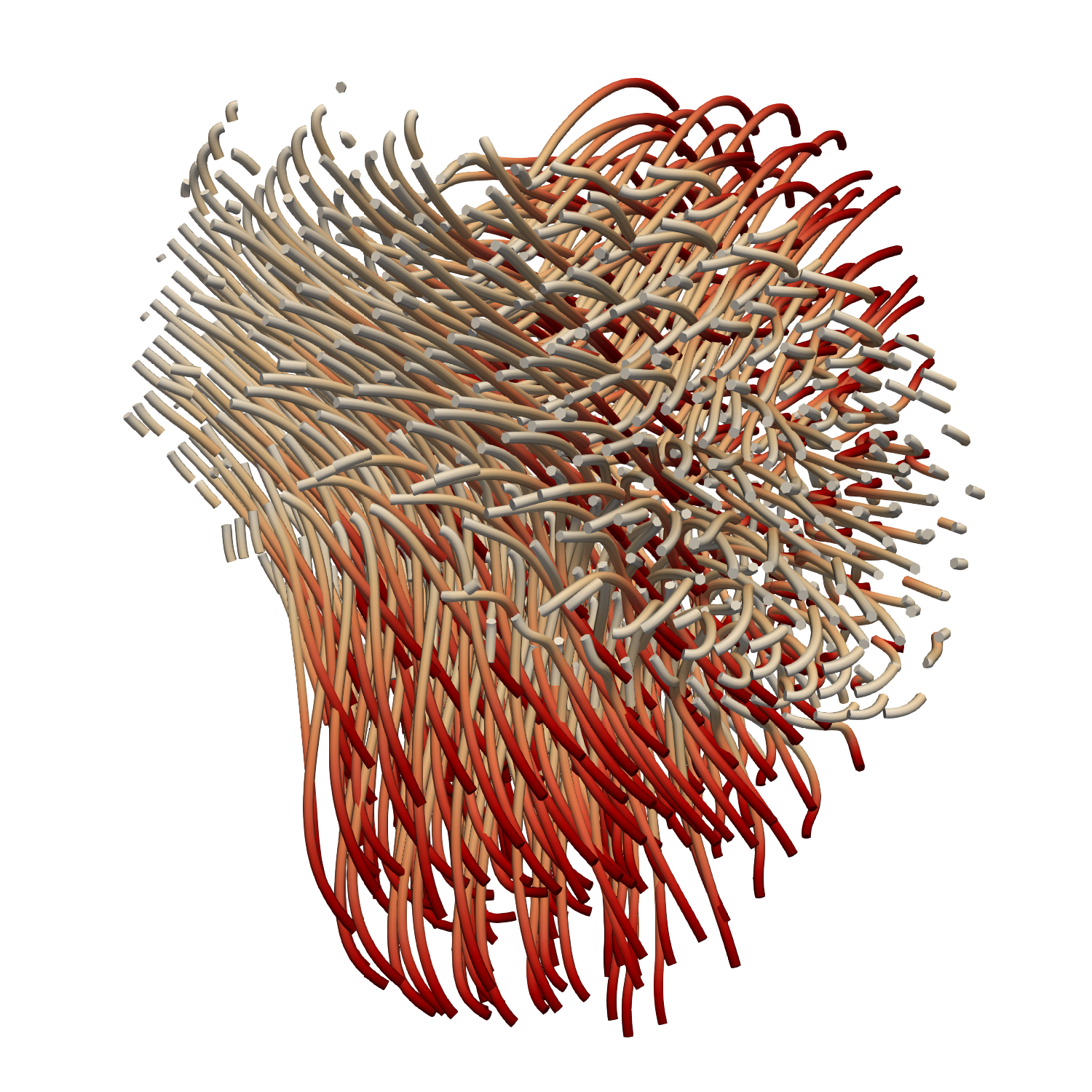}
      \label{fig:dropcol-results-separation-path-500-orig}}%
    \hfill%
    \subfigure[Path]{\includegraphics[width=\imagewidth\linewidth,trim=0 \clipTB cm 0 \clipTB cm,clip]{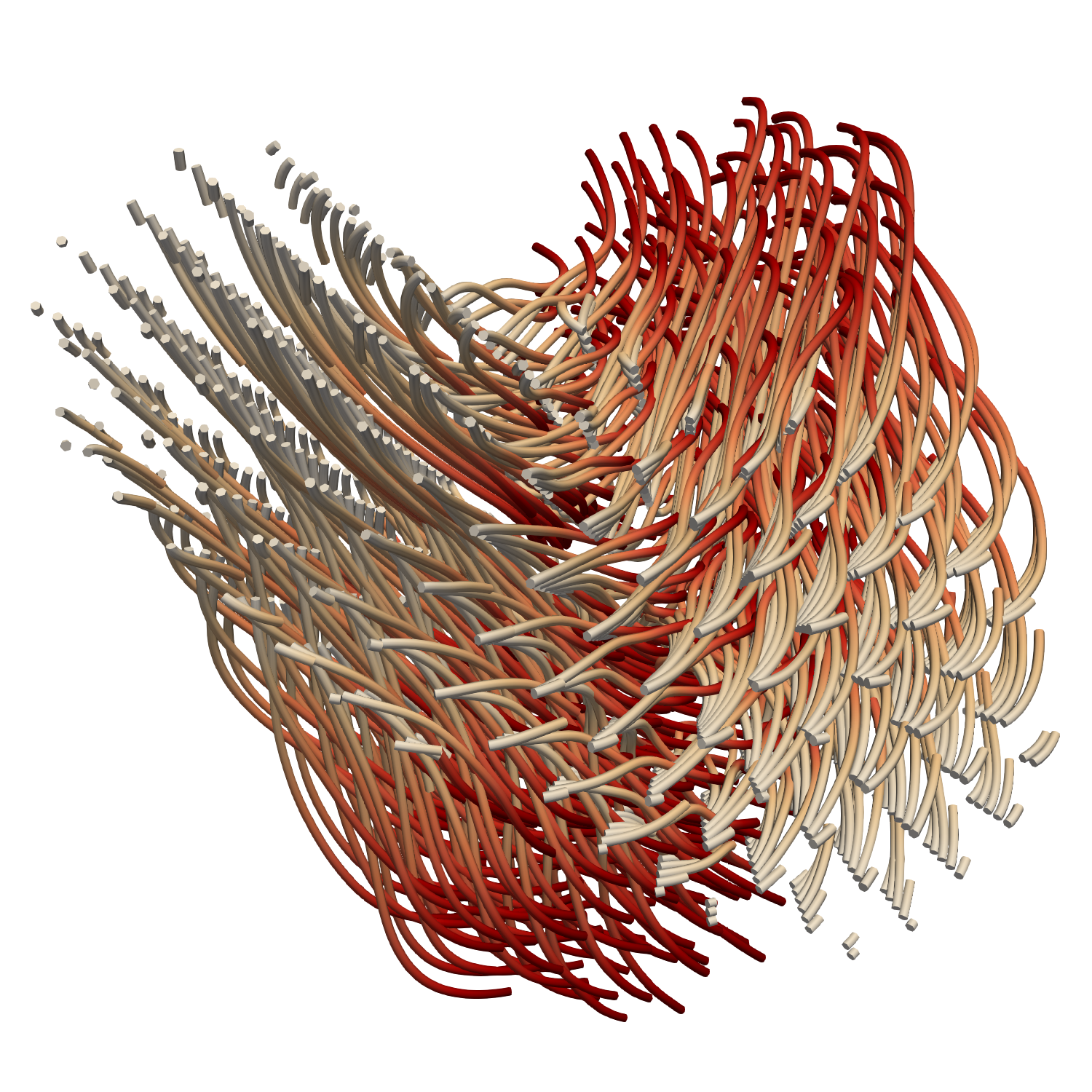}
      \label{fig:dropcol-results-separation-path-785-orig}}%
    \hfill%
    \subfigure[Streak]{\includegraphics[width=\imagewidth\linewidth,trim=0 \clipTB cm 0 \clipTB cm,clip]{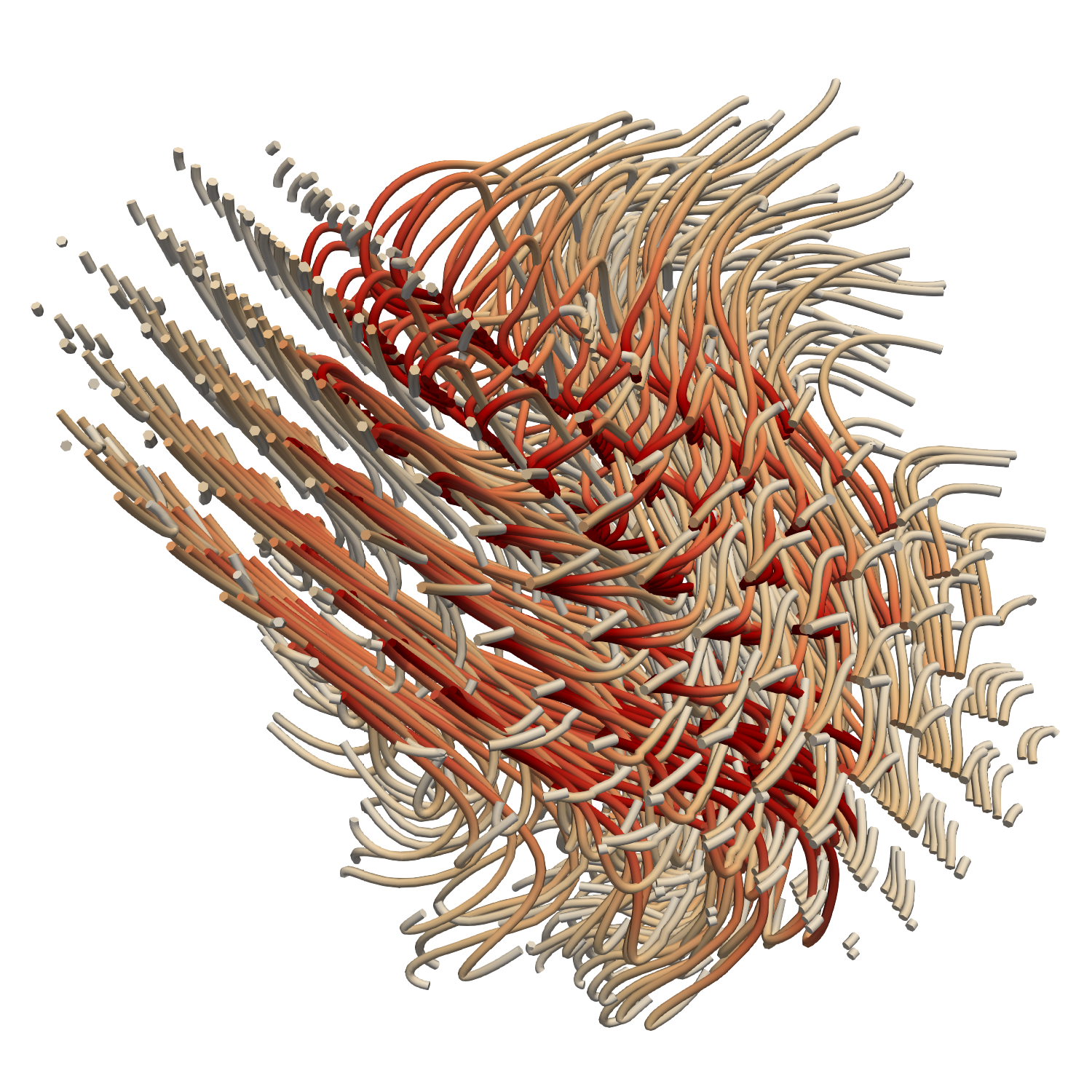}
      \label{fig:dropcol-results-separation-streak-785-orig}}%
    \hfill%
    \\[0.5em]%
    \hfill%
    \subfigure[Stream]{\includegraphics[width=\imagewidth\linewidth,trim=0 \clipTB cm 0 0,clip]{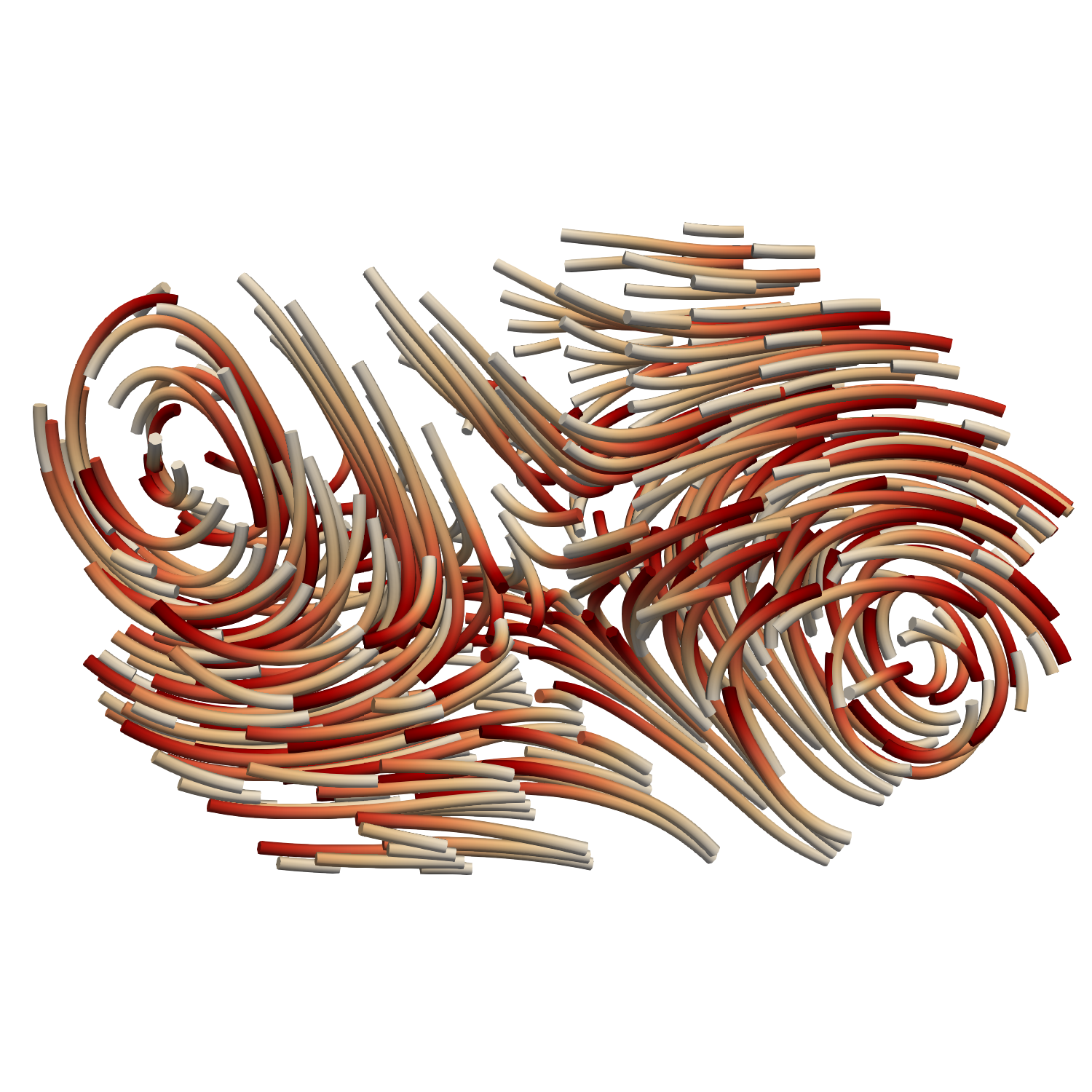}
      \label{fig:dropcol-results-vortex-stream-500}}%
    \hfill%
    \subfigure[Path]{\includegraphics[width=\imagewidth\linewidth,trim=0 \clipTB cm 0 0,clip]{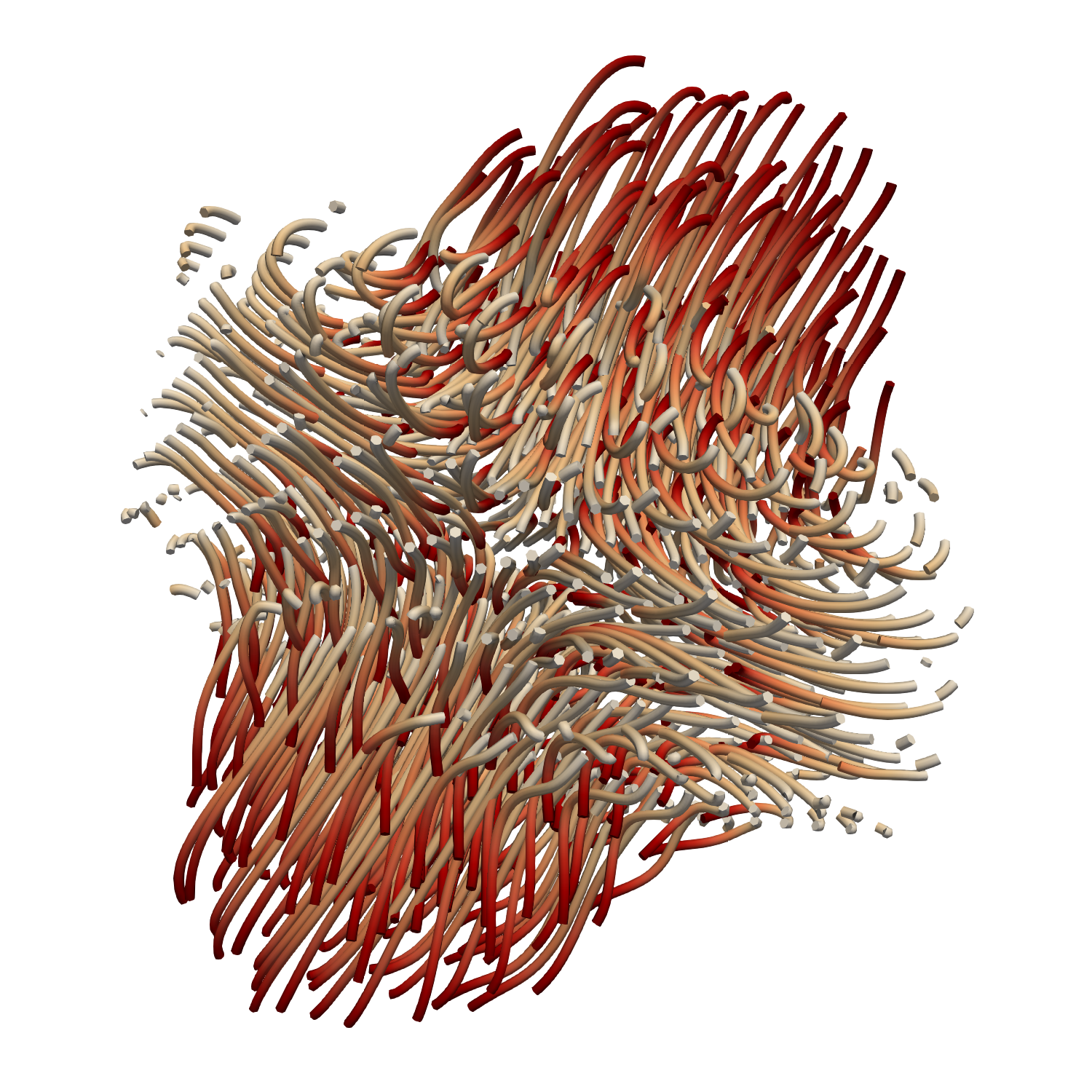}
      \label{fig:dropcol-results-separation-path-500}}%
    \hfill%
    \subfigure[Path]{\includegraphics[width=\imagewidth\linewidth,trim=0 \clipTB cm 0 0,clip]{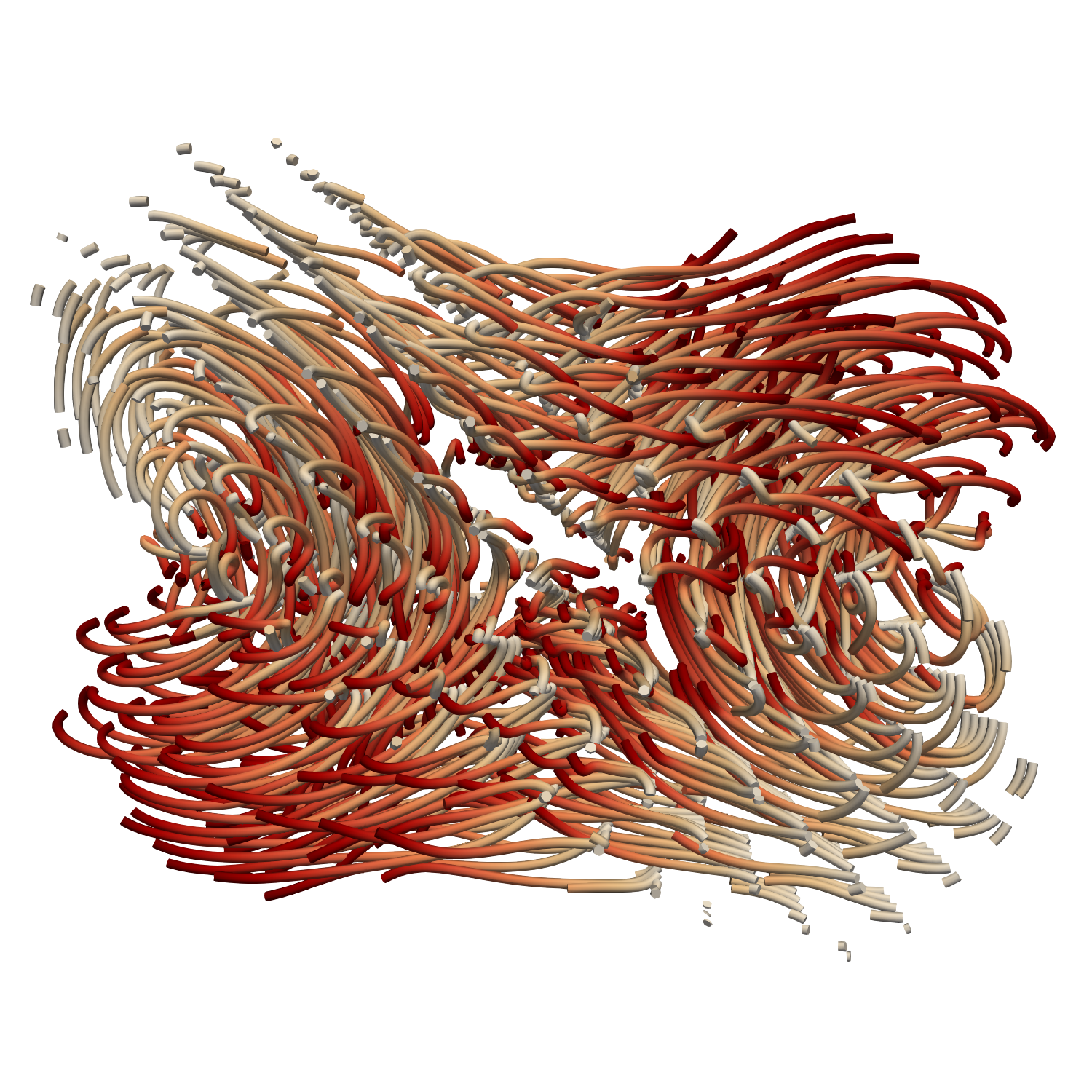}
      \label{fig:dropcol-results-separation-path-785}}%
    \hfill%
    \subfigure[Streak]{\includegraphics[width=\imagewidth\linewidth,trim=0 \clipTB cm 0 0,clip]{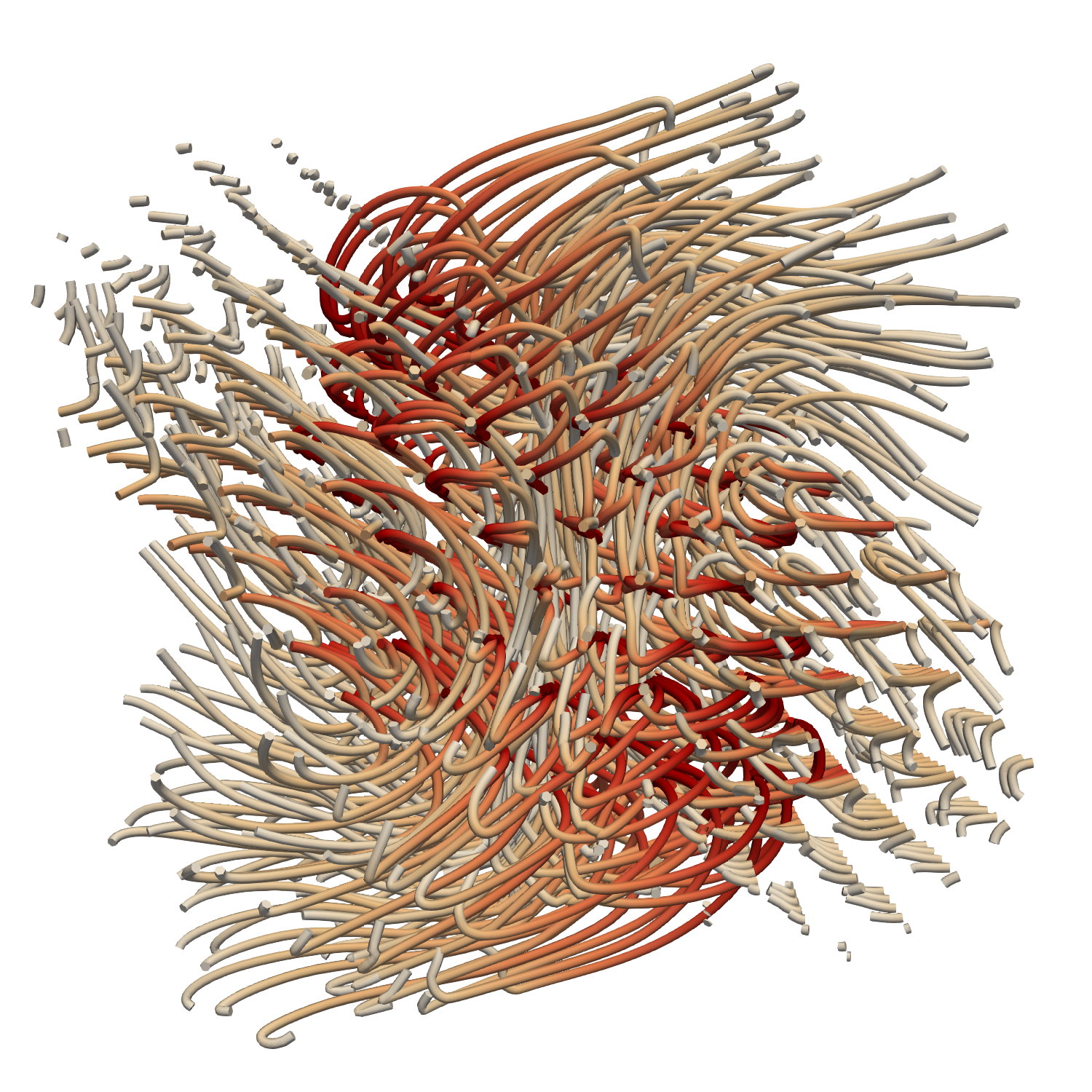}
      \label{fig:dropcol-results-separation-streak-785}}%
    \hfill%
  \end{minipage}%
  \caption{%
    Examples from different time steps of the~\dropcol\ data\-set in columns.
    \protect\subref{fig:dropcol-results-vortex-stream-500-orig}--\protect\subref{fig:dropcol-results-separation-streak-785-orig}~Original and \protect\subref{fig:dropcol-results-vortex-stream-500}--\protect\subref{fig:dropcol-results-separation-streak-785}~droplet-local stream-, path-, and streaklines.
    Flow direction is encoded from white to red.
  }
  \label{fig:dropcol-results}
\end{figure}

Especially particles seeded in static vector fields exhibiting large translational velocity tend to quickly leave their respective phase~(\autoref{fig:dropcol-results-vortex-stream-500-orig}).
Thus, streamlines are often unsuitable and only show the general movement of the droplet.
However, their droplet-local counterparts may reveal intricate details of droplet-internal flow~(\autoref{fig:dropcol-results-vortex-stream-500}).
Here, a saddle-like structure can be observed in the center of the droplet, while to the sides, there are two vortices.
But, also pathlines can reveal more details by employing a droplet-local frame of reference.
While for the original vector field they show the actual flow~(\autoref{fig:dropcol-results-separation-path-500-orig}), applying them to our droplet-local method much better reveals the separating behavior of the lower left and the upper right region of the droplet~(\autoref{fig:dropcol-results-separation-path-500}).
Specifically, it can be observed that there exists a plane in between these two regions that separates flow from going into the upward or the downward part of the droplet.
Similar behavior is apparent in the example in~\autoref{fig:dropcol-results-separation-path-785}.
Again, a separation can be observed, but this time with a more vortical motion in the lower left and upper right parts.
This observation can barely be made from the original pathlines~(\autoref{fig:dropcol-results-separation-path-785-orig}).
Streaklines for this example are additionally visualized in~\autoref{fig:dropcol-results-separation-streak-785-orig} and~\autoref{fig:dropcol-results-separation-streak-785}.

\section{Discussion}
\label{sec:discussion}

A particularity of multiphase flow is that particles, although fictitious, belong to a certain fluid phase.
These particles, therefore, have to stay within the same phase in the course of the simulation.
Additionally, VOF fields cannot be interpolated between time steps in a way that droplet interfaces are correctly preserved.
This \mbox{makes} it difficult to apply higher-order integration schemes, requiring interpolation between phases at droplet boundaries.
Therefore, we employ Adams--Bashforth integration for streak- and pathline computation, as well as Runge--Kutta 4 for streamline integration.
Still, particles may leave the droplet due to numerical inaccuracy.
Then, the integration of an integral line has to be stopped, as the local velocity field is only defined for cells within the respective droplet.

Because our method allows the use of a static or a dynamic frame of reference per droplet, i.e., the frame of reference is either extracted from the initial droplet at the start of the integration or the frame of reference is extracted anew for every time step, we can apply our approach to different scenarios.
While a static frame of reference might be easier to interpret, a dynamic frame of reference can, for example, be used on datasets where the angular velocity increases or decreases.
This is, e.g., the case in stellar mergers where two stars rotate around a common center of mass until they finally merge into a single star.
Here, the rotation around this common center changes over time.
To allow domain scientists to observe the mass transfer between the stars, it is useful to keep them at their respective positions by applying a dynamic frame of reference.

Co-rotating grids are usually implemented by resampling the grid.
For multiple frames of reference, such as droplet-local frames, re\-samp\-ling would need to be performed for each droplet se\-pa\-rate\-ly.
A topological event at a time~$t_e$, e.g., the collision of droplets, would then have to be handled explicitly by duplicating a subgrid at~$t_e$ for two different sources (the two colliding droplets) at their original positions at start time~$t_0$ of the integration.
Additionally, segmentation into droplets might need to be performed on a scalar field that cannot be resampled, as is the case for a VOF field.
On the contrary, our method does not have to perform resampling, and topological changes of the droplets, i.e., collisions and breakups, do not have to be handled explicitly.
Droplet tracking from our context visualization approach can, however, detect such events and gives feedback to the user.
Using a static frame of reference, breakups can easily be interpreted, as the flow will naturally show this separation process.
When dealing with collisions, an abrupt change in direction and velocity magnitude occurs.
However, information about how the impact occurred is missing.
As collisions are not predictable from observing droplets separately, this is a difficult challenge.
While the static frame of reference breaks down in such a case, our dynamic frame of reference yields good results~(see supplemental video).

Performance-wise, the most costly part of our approach is data loading and droplet extraction.
Compared to the calculation of traditional integral lines, the cost for advection is doubled, as now two lines are integrated simultaneously.
To cancel out the rotation, the sampled droplet-local velocities additionally have to be transformed.
However, if we were to resample the grid to give us a droplet-local velocity field, this would also entail the previously mentioned steps for pre-computation.
Compared to resampling, our method should be significantly faster when dealing with a number of streamlines that is much smaller than the number of grid cells.
When dealing with droplets, this is typically the case.

\section{Conclusion}
\label{sec:conclusion}

In this paper, we have described and shown the utility of vi\-sualization of droplet-local velocity fields in three-dimensional multi\-phase flow.
To this end, we presented the calculation of said derived velocity field and introduced a modified, generalized ap\-proach for pathline and streakline computation.
Additionally, removed information is visualized using glyphs for context.
We applied our technique to different droplets, showing the usefulness of our method: droplet-local line integration yields interesting results for most small and medium-sized droplets that exhibit individual, characteristic behavior overshadowed by relatively high rotational or translational velocity.
In the case of nearly static droplets, a challenging idea could be to use clustering to visualize ``cluster-local velocity''.
This would lead to a more generalized approach, which would allow the separate inspection of clusters and, therefore, the analysis of larger and more complex fluid structures.
For future work, an idea is to integrate our method into an in situ environment.
Further, our method can readily be combined with seeding strategies and opacity optimization to reduce visual clutter.
For a different application, the droplet-local velocity field could also be used to provide more numerical stability to feature extraction, e.g., extraction of vortices in droplets.

\acknowledgments{%
This work was partially funded by Deutsche Forschungsgemeinschaft~(DFG) as part of the Cluster of Excellence EXC 2075 ``Sim\-Tech''~(390740016), International Research Training Group GRK 2160 ``DROPIT''~(270852890) and Collaborative Research Centre SFB 1313~(327154368) at the University of Stuttgart, and as part of the Transregional Collaborative Research Centre SFB/TRR 165 ``Waves to Weather''~(257899354) at Heidelberg University.}

\bibliographystyle{abbrv-doi}
\bibliography{paper}

\begin{thebibliography}{10}

\bibitem{DROPIT}
International research training group 2160: {D}roplet interaction technologies
  ({DROPIT}).
\newblock https://www.project.uni-stuttgart.de/dropit/.

\bibitem{Ayachit2015}
U.~Ayachit.
\newblock {\em The {ParaView} Guide: {A} Parallel Visualization Application}.
\newblock Kitware, Inc., 2015.

\bibitem{Bhatia2014}
H.~Bhatia, V.~Pascucci, and P.-T. Bremer.
\newblock The natural {H}elmholtz-{H}odge decomposition for open-boundary flow
  analysis.
\newblock {\em {IEEE} Transactions on Visualization and Computer Graphics},
  20(11):1566--1578, 2014.

\bibitem{Brambilla2013}
A.~Brambilla, {\O}.~Andreassen, and H.~Hauser.
\newblock Integrated multi-aspect visualization of {3D} fluid flows.
\newblock In {\em Vision, Modeling and Visualization, {VMV} 2013}, pp. 1--9,
  2013.

\bibitem{Eisenschmidt2016}
K.~Eisenschmidt, M.~Ertl, H.~Gomaa, C.~Kieffer-Roth, C.~Meister,
  P.~Rauschenberger, M.~Reitzle, K.~Schlottke, and B.~Weigand.
\newblock Direct numerical simulations for multiphase flows: An overview of the
  multi\-phase code {FS3D}.
\newblock {\em Applied Mathematics and Computation}, 272:508--517, 2016.

\bibitem{Guenther2017a}
T.~G{\"{u}}nther, M.~H. Gross, and H.~Theisel.
\newblock Generic objective vortices for flow visualization.
\newblock {\em {ACM} Transactions on Graphics}, 36(4):141:1--141:11, 2017.

\bibitem{Guenther2014}
T.~G{\"{u}}nther, C.~R{\"{o}}ssl, and H.~Theisel.
\newblock Hierarchical opacity optimization for sets of {3D} line fields.
\newblock {\em Computer Graphics Forum}, 33(2):507--516, 2014.

\bibitem{Guenther2017}
T.~G{\"{u}}nther, H.~Theisel, and M.~H. Gross.
\newblock Decoupled opacity optimization for points, lines and surfaces.
\newblock {\em Computer Graphics Forum}, 36(2):153--162, 2017.

\bibitem{Hadwiger2019}
M.~Hadwiger, M.~Mlejnek, T.~Theu{\ss}l, and P.~Rautek.
\newblock Time-dependent flow seen through approximate observer {Killing}
  fields.
\newblock {\em {IEEE} Trans\-ac\-tions on Visualization and Computer Graphics},
  25(1):1257--1266, 2019.

\bibitem{Hirt1981}
C.~W. Hirt and B.~D. Nichols.
\newblock Volume of fluid ({VOF}) method for the dynamics of free boundaries.
\newblock {\em Journal of Computational Physics}, 39(1):201--225, 1981.

\bibitem{Hlawatsch2014}
M.~Hlawatsch, F.~Sadlo, H.~Jang, and D.~Weiskopf.
\newblock Pathline glyphs.
\newblock {\em Computer Graphics Forum}, 33(2):497--506, 2014.

\bibitem{Karch2018}
G.~K. Karch, F.~Beck, M.~Ertl, C.~Meister, K.~Schulte, B.~Weigand, T.~Ertl, and
  F.~Sadlo.
\newblock Visual analysis of inclusion dynamics in two-phase flow.
\newblock {\em {IEEE} Transactions on Visualization and Computer Graphics},
  24(5):1841--1855, 2018.

\bibitem{Li2006}
H.~Li, W.~Chen, and I.~Shen.
\newblock Segmentation of discrete vector fields.
\newblock {\em {IEEE} Transactions on Visualization and Computer Graphics},
  12(3):289--300, 2006.

\bibitem{Marchesin2010}
S.~Marchesin, C.~Chen, C.~Ho, and K.~Ma.
\newblock View-dependent streamlines for {3D} vector fields.
\newblock {\em {IEEE} Transactions on Visualization and Computer Graphics},
  16(6):1578--1586, 2010.

\bibitem{McLoughlin2013}
T.~McLoughlin, M.~W. Jones, R.~S. Laramee, R.~Malki, I.~Masters, and C.~D.
  Hansen.
\newblock Similarity measures for enhancing interactive streamline seeding.
\newblock {\em {IEEE} Transactions on Visualization and Computer Graphics},
  19(8):1342--1353, 2013.

\bibitem{Straub2020}
A.~Straub and T.~Ertl.
\newblock Visualization techniques for droplet interfaces and multiphase flow.
\newblock In {\em Droplet Interactions and Spray Processes}, vol. 121 of {\em
  Fluid Mechanics and its Applications}, pp. 203--214. 2020.

\bibitem{Straub2019}
A.~Straub, M.~Heinemann, and T.~Ertl.
\newblock Visualization and visual analysis for multiphase flow.
\newblock In {\em Proceedings of the DIPSI Workshop 2019}, pp. 25--27, 2019.

\bibitem{Turk1996}
G.~Turk and D.~Banks.
\newblock Image-guided streamline placement.
\newblock In {\em Proceedings of {SIGGRAPH} 96}, pp. 453--460, 1996.

\bibitem{Wiebel2005}
A.~Wiebel, C.~Garth, and G.~Scheuermann.
\newblock Localized flow analysis of {2D} and {3D} vector fields.
\newblock In {\em Joint Eurographics -- {IEEE} {VGTC} Symposium on
  Visualization, {EuroVis} 2005}, pp. 143--150, 2005.

\bibitem{Wiebel2007}
A.~Wiebel, X.~Tricoche, D.~Schneider, H.~J{\"{a}}nicke, and G.~Scheuermann.
\newblock Generalized streak lines: Analysis and visualization of boundary
  induced vortices.
\newblock {\em {IEEE} Transactions on Visualization and Computer Graphics},
  13(6):1735--1742, 2007.

\end{thebibliography}
\end{document}